# Programing implementation of the Quine-McCluskey method for minimization of Boolean expression

Jiangbo Huang
Department of Biological Sciences, Faculty of Science
National University of Singapore, Singapore 117604

## Abstract

A Boolean function is a function that produces a Boolean value output by logical calculation of Boolean inputs. It plays key roles in programing algorithms and design of circuits. Minimization of Boolean function is able to optimize the algorithms and circuits. Quine-McCluskey (QM) method is one of the most powerful techniques to simplify Boolean expressions. Compared to other techniques, QM method is more executable and can handle more variables. In addition, QM method is easier to be implemented in computer programs, which makes it an efficient technique. There are several versions of QM simulation codes online, whereas some of them appear to have limitations of variables numbers or lack the consideration of Don't-Care conditions. Here a QM simulation code based on C programing is introduced. Theoretically it is able to handle any number of variables and has taken the Don't-Care conditions into account.

## 1. Introduction

Simplification of Boolean expression is a practical tool to optimize programing algorithms and circuits. Several techniques have been introduced to perform the minimization, including Boolean algebra (BA), Karnaugh Map (K-Map) and QM. Minimization using BA requires high algebraic manipulation skills and will become more and more complicated when the number of terms increases. K-Map is a diagrammatic technique based on a special form of Venn diagram. It is easier to use than BA but usually it is used to handle Boolean expression with no more than six variables. When the number of variables exceeds six, the complexity of the map is exponentially enhanced and it becomes more and more cumbersome. Functionally identical to K-Map, QM method is more executable when dealing with larger number of variables and is easier to be mechanized and run on a computer. Although a number of programing codes implementing QM method are available online, not all of them are technically correct. Furthermore, it is found that some of them either didn't take the Don't-Care conditions into consideration or still had limitation of the number of variables. Here a QM simulation code based on C language is introduced. Theoretically speaking, it has no limitation of the number of variables and has taken the Don't-Care conditions into account. In this article, we will introduce the procedure of QM method step by step. In each step we will explain how we implement it in the program. At last we will attach the codes.

## 2. Procedures and implementation of QM method

2.1 General introduction of the procedures

The QM method is based on the reduction principle, which says that

$$AB+A\overline{B} = A$$

In this formula, A can be either a variable or a team and B is a variable. It means that when two terms contain the same variables differ only in one variable, they can be combined together and form a new term smaller by one literal. All the terms in the Boolean function are tested for possible combination of any two of them, after which a new sets of terms that are smaller by one literal are produced and are further tested under the same procedures for further reduction. The same procedures will be repeated until no terms can be combined anymore. The irreducible terms are named 'Prime Implicant' (PI). The final step is to select a sets of PIs which contain the least possible number of PIs and cover all the original terms. The selected PIs are called 'Essential Prime Implicant' (EPI). The EPIs represent the final minimized expression.

2.2 Procedure of QM method and algorithm for implementation

To illustrate the detailed procedures of QM methods, let's consider the following example of Boolean function:

$$F(A,B,C,D) = \sum m(4,5,6,9,11,12,13,14) + \sum d(0,1,3,7)$$

Which totally contain 12 minterms, including 4 Don't-Care minterms.

| Procedures of QM method | Data structure and algorithm for implementation |
|---|---|
| 1. All the minterms are transformed into binary numbers and arranged into different groups according to the number of 1s in the binary representation. These groups are put in ascending order in the Column 0 (Fig 1). | 1. The decimal indices will be translated into binary formats and stored in the two-dimensional array Minterm_Binary[][]. In each single-dimensional array Minterm_Binary[*i*][], the first four members are the corresponding binary digits; the fifth member indicates the number of 1s; the sixth member and the seventh member are the corresponding decimal number (Fig 2). After grouping, they will be put into the array Column[0][*i*][], where 0 means they are from column 0 and *i* indicates that the number of 1s is *i* (Fig 3); |
| 2. Every two minterms from two adjacent groups are paired for possible combination if they only differ in one variable. The digit being canceled is replaced by 'X'. After the first round of pairing, a new sets of terms that are smaller by one literal are produced and placed in column 1. Terms in Column 1 are further paired for further reduction. The same procedures will be repeated until no terms can be paired anymore. In these columns, all the terms that has been paired will be ticked (Fig 4). | 2. To store the terms in the columns, the four-dimensional array Column[][][][] is introduced. Each single-dimensional array Column[*i*][*j*][*k*][] is produced dynamically, where *i* indicates the number of column; *j* is the number of group; *k* is the number of each term in each group. The size and members of array Column[*i*][*j*][*k*][] is illustrated in Fig 5. In the first step of pairing term Column[*i*][*j*][*k*][] and term Column[*i*][*j*][*l*][], it will be checked whether the elements from Column[*i*][*j*][*k*][6] to Column[*i*][*j*][*k*][5+*i*] are the same as the elements from Column[*i*][*j*+1][*l*][6] to Column[*i*][*j*+1][*l*][5+*i*] (Which indicates whether term Column[*i*][*j*][*k*][] and term Column[*i*][*j*+1][*l*][] have the same position of 'X's). The next condition to check is whether the difference between Column[*i*][*j*][*k*][6+*i*] and Column [*i*][*j*+1][*l*][6+*i*] is power of two or not (which is equivalent to whether term Column[*i*][*j*][*k*][] and term Column[*i*][*j*+1][*l*][] only differ in one variable). If both these two conditions are true, it means that |
| | the two terms can be paired. Then the array Column[$i$+1][$j$][$m$][] will be produced to store the information of the combined term. The value of Column[$i$][$j$][$k$][5] and Column[$i$][$j$+1][$l$][5] will be changed from 0 to 1, which means they are ticked (Fig 6). |
| 3. Notably, terms with more than one 'X' will always have duplicates. For example, in column 2, the terms 0X0X can be obtained from 000X and 010X or 0X00 and 0X01. Thus, all the duplicates need to be removed (Fig 7). The remaining irreducible terms (which are not ticked) are PIs. | 3. After finishing all possible pairings, the address information of unpaired terms (Column[$i$][$j$][$k$][5] is 0) will be stored in the array PI_Index[$n$][] and duplicates will be ignored. For example, if Column[2][1][9] is a PI, the elements of the corresponding PI_Index[$n$][] will be {2,1,9} (Fig 8).<br>In the same time, an array named NumberCounter[] will be produced to record how many times every decimal index occurs in all the PIs. |
| 4. A 'PI chart' will be made as shown in Fig 9. Each column represents a minterm and each row represents a PI. In the first step, all the Don't-Care minterms is removed (Fig 9). | 4. All the Don't-Care minterms will be removed by simply change the value of the corresponding NumberCounter[$i$] to 0 (Fig 10). |
| 5. Column with only a single red circle indicates a minterm covered by only one PI, which means this PI is EPI. All the rows of EPI and the corresponding columns that are covered will be removed (Fig 11). | 5. The program will go through the array NumberCounter[].When NumberCounter[$k$] is 1, the PI that contains this minterm $k$ will be selected and the information of the corresponding PI_Index[][] will be transferred to EPI_Index[][].<br>The value of the corresponding NumberCounter[$n$] of the minterms covered by these EPIs will be set to 0 (Fig 12). |
| 6. The remaining PIs and minterms will form a new | 6. The Reduced PI chart will be produced as follows: the number of columns is denoted by |
| 'Reduced PI chart. The next step is to find a sets of PIs which cover all the terms in the Reduced PI chart but contain the least possible number of PIs. This set of PIs will be the rest EPIs. Obviously, either XX01 or X10X can be the last EPI (Fig 13). | NumberOfRemainingMT; the number of rows is denoted by NumberOfRemainingPI; The remaining minterms are recorded in ReducedPIChart_X[]; the remaining PIs are recorded in ReducedPIChart_Y[][]; the two-dimentional array ReducedPIChart[][] will denote the red circles (Fig 14).<br><br>In the first step, the program will find all the possible sets of PIs that cover all the remaining minterms using nested for loop. The number of layers of the nested for loop depends on the number of the remaining minterms. However, the number of the remaining minterms depends on the input of the program, which is unknown when writing the program. To solve this problem, a recursive function is produced. After going through all the possible sets of PIs, a simple for loop will be utilized to find the set containing the least number of PIs. The PIs in this set will be the rest EPIs (Fig 14). |

## Procedures of QM method (1)

| Decimal | Binary |
|---|---|
| 0 | 0000 |
| 1 | 0001 |
| 3 | 0011 |
| 4 | 0100 |
| 5 | 0101 |
| 6 | 0110 |
| 7 | 0111 |
| 9 | 1001 |
| 11 | 1011 |
| 12 | 1100 |
| 13 | 1101 |
| 14 | 1110 |

| Column 0 | |
|---|---|
| 0 | 0000 |
| 1 | 0001 |
| 4 | 0100 |
| 3 | 0011 |
| 5 | 0101 |
| 6 | 0110 |
| 9 | 1001 |
| 12 | 1100 |
| 7 | 0111 |
| 11 | 1011 |
| 13 | 1101 |
| 14 | 1110 |

Figure 1. Binary representation of the minterms and grouping according to the number of 1s

## Data structure and algorithm for implementation (1)

NumberOfVariable=4;

NumberOfAllMinterm=12;

NumberOfDontCare=4;

MintermIndicesDecimal[]={0,1,3,4,5,6,7,9,11,12,13,14};

MintermIndicesDecimal_DontCare[]={0,1,3,7}

Minterm_Binary[*i*][*j*]

| *i* \ *j* | 0 | 1 | 2 | 3 | 4 | 5 | 6 | 7 |
|---|---|---|---|---|---|---|---|---|
| 0 | 0 | 0 | 0 | 0 | 0 | 0 | 0 | 0 |
| 1 | 0 | 0 | 0 | 1 | 1 | 0 | 1 | 1 |
| 2 | 0 | 0 | 1 | 1 | 2 | 0 | 3 | 3 |
| 3 | 0 | 1 | 0 | 0 | 1 | 0 | 4 | 4 |
| 4 | 0 | 1 | 0 | 1 | 2 | 0 | 5 | 5 |
| 5 | 0 | 1 | 1 | 0 | 2 | 0 | 6 | 6 |
| 6 | 0 | 1 | 1 | 1 | 3 | 0 | 7 | 7 |
| 7 | 1 | 0 | 0 | 1 | 2 | 0 | 9 | 9 |
| 8 | 1 | 0 | 1 | 1 | 3 | 0 | 11 | 11 |
| 9 | 1 | 1 | 0 | 0 | 2 | 0 | 12 | 12 |
| 10 | 1 | 1 | 0 | 1 | 3 | 0 | 13 | 13 |
| 11 | 1 | 1 | 1 | 0 | 3 | 0 | 14 | 14 |

Figure 2. The information of binary numbers are stored in the array Minterm_Binary[][]

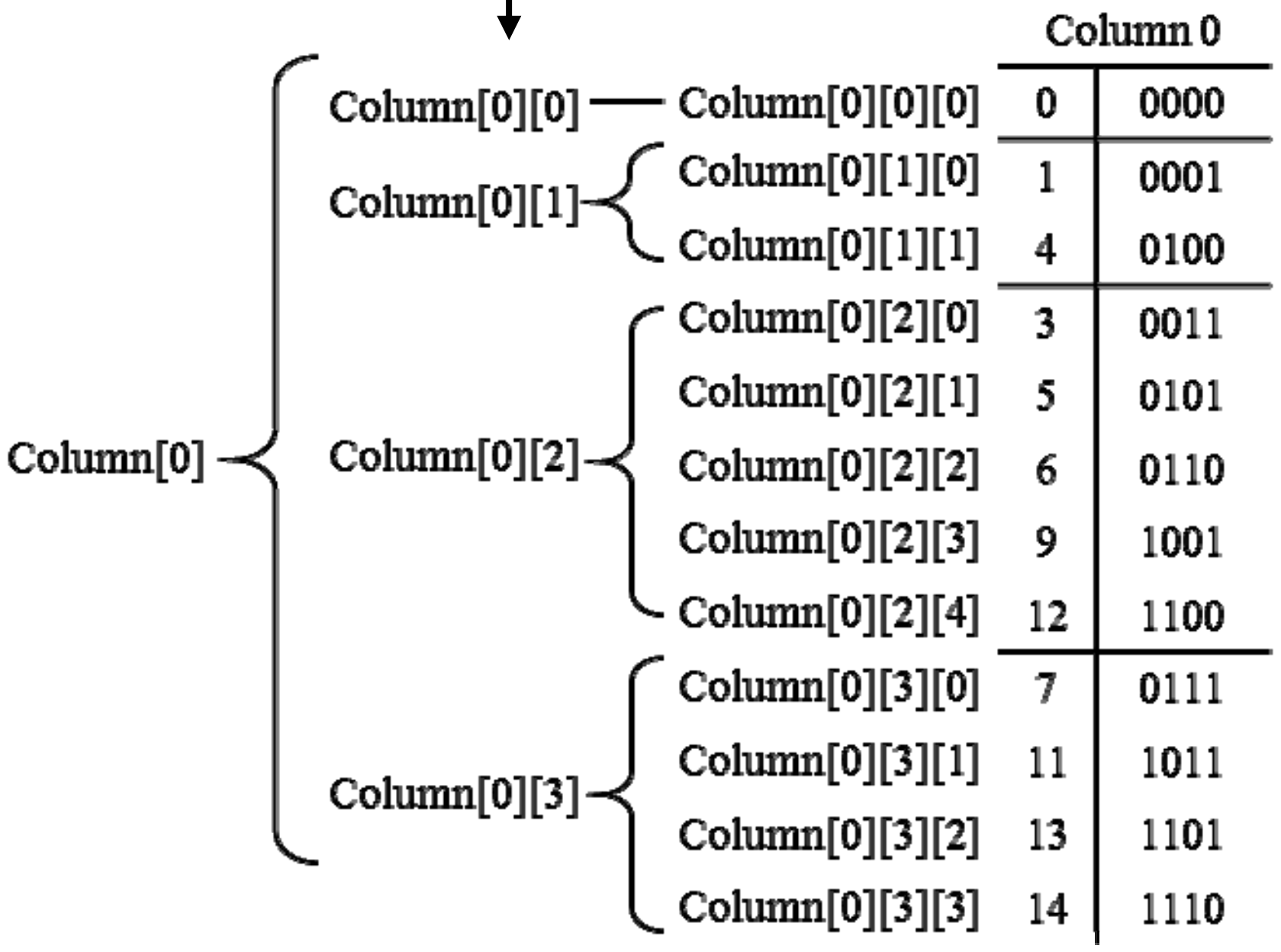

Figure 3. Arrange the minterms into groups according to the number of 1s in their binary representation

## Procedures of QM method (2)

Column 0

| | | |
|---|---|---|
| 0 | 0000 | ✓ |
| 1 | 0001 | ✓ |
| 4 | 0100 | ✓ |
| 3 | 0011 | ✓ |
| 5 | 0101 | ✓ |
| 6 | 0110 | ✓ |
| 9 | 1001 | ✓ |
| 12 | 1100 | ✓ |
| 7 | 0111 | ✓ |
| 11 | 1011 | ✓ |
| 13 | 1101 | ✓ |
| 14 | 1110 | ✓ |

Column 1

| | |
|---|---|
| 0,1 | 000X |
| 0,4 | 0X00 |
| 1,3 | 00X1 |
| 1,5 | 0X01 |
| 1,9 | X001 |
| 4,5 | 010X |
| 4,6 | 01X0 |
| 4,12 | X100 |
| 3,7 | 0X11 |
| 3,11 | X011 |
| 5,7 | 01X1 |
| 5,13 | X101 |
| 6,7 | 011X |
| 6,14 | X110 |
| 9,11 | 10X1 |
| 9,13 | 1X01 |
| 12,13 | 110X |
| 12,14 | 11X0 |

Fig 4. First round of pairing from Column 0 to Column 1

## Data structure and algorithm for implementation (2)

Members of Column[*i*][*j*][*k*][]

(Size of Column[*i*][*j*][*k*] is $7+i+2^i$)

| *l* | Column[*i*][*j*][*k*][*l*] |
|---|---|
| 0<br>1<br>2<br>3 | Binary representation of indices. Positions of X are replaced by 2 |
| 4 | Number of 1s |
| 5 | ‘1’ means have been paired off; ‘0’ means haven’t been paired off |
| 6 …… (5+*i*) | Positions of the ‘X’s |
| 6+*i* | The original decimal index |
| (7+*i*) …… ($6+i+2^i$) | All the decimal indices included |

Fig 5 Members of array Column[*i*][*j*][*k*][]

| *l* | 0 | 1 | 2 | 3 | 4 | 5 | 6 | 7 | 8 | 9 |
|---|---|---|---|---|---|---|---|---|---|---|
| Column[1][1][4][*l*] | 0 | 1 | 2 | 0 | 1 | 0 | 1 | 4 | 4 | 6 |

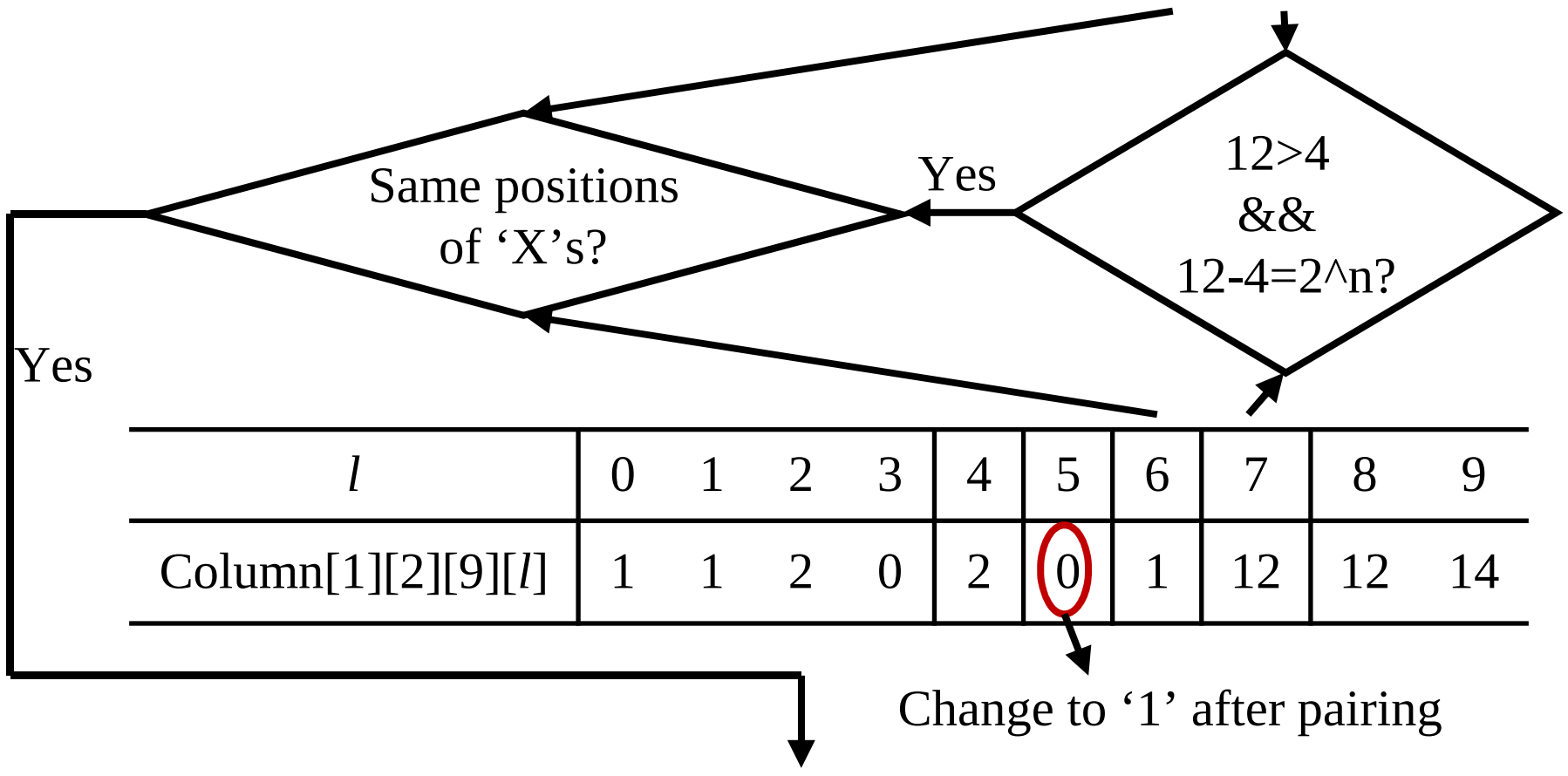

| *l* | 0 | 1 | 2 | 3 | 4 | 5 | 6 | 7 | 8 | 9 |
|---|---|---|---|---|---|---|---|---|---|---|
| Column[1][2][9][*l*] | 1 | 1 | 2 | 0 | 2 | 0 | 1 | 12 | 12 | 14 |

| *l* | 0 | 1 | 2 | 3 | 4 | 5 | 6 | 7 | 8 | 9 | 10 | 11 | 12 |
|---|---|---|---|---|---|---|---|---|---|---|---|---|---|
| Column[2][1][9][*l*] | 2 | 1 | 2 | 0 | 1 | 0 | 1 | 3 | 4 | 4 | 6 | 12 | 14 |

Fig 6 Implementation of pairing process

# Procedures of QM method (3)

Column 1

| | | |
|---|---|---|
| 0,1 | 000X | ✓ |
| 0,4 | 0X00 | ✓ |
| 1,3 | 00X1 | ✓ |
| 1,5 | 0X01 | ✓ |
| 1,9 | X001 | ✓ |
| 4,5 | 010X | ✓ |
| 4,6 | 01X0 | ✓ |
| 4,12 | X100 | ✓ |
| 3,7 | 0X11 | ✓ |
| 3,11 | X011 | ✓ |
| 5,7 | 01X1 | ✓ |
| 5,13 | X101 | ✓ |
| 6,7 | 011X | ✓ |
| 6,14 | X110 | ✓ |
| 9,11 | 10X1 | ✓ |
| 9,13 | 1X01 | ✓ |
| 12,13 | 110X | ✓ |
| 12,14 | 11X0 | ✓ |

→

Column 2

| | |
|---|---|
| 0,1,4,5 | 0X0X |
| ~~0,4,1,5~~ | ~~0X0X~~ |
| 1,3,5,7 | 0XX1 |
| 1,3,9,11 | X0X1 |
| ~~1,5,3,7~~ | ~~0XX1~~ |
| 1,5,9,13 | XX01 |
| ~~1,9,3,11~~ | ~~X0X1~~ |
| ~~1,9,5,13~~ | ~~XX01~~ |
| 4,5,6,7 | 01XX |
| 4,5,12,13 | X10X |
| ~~4,6,5,7~~ | ~~01XX~~ |
| 4,6,12,14 | X1X0 |
| ~~4,12,5,13~~ | ~~X10X~~ |
| ~~4,12,6,14~~ | ~~X1X0~~ |

Fig 7 The last round of pairing and omitting the duplicate

---

# Data structure and algorithm for implementation (3)

| *l* | 0 | 1 | 2 | 3 | 4 | 5 | 6 | 7 | 8 | 9 | 10 | 11 | 12 |
|---|---|---|---|---|---|---|---|---|---|---|---|---|---|
| Column[2][1][9][*l*] | 2 | 1 | 2 | 0 | 1 | 0 | 1 | 3 | 4 | 4 | 6 | 12 | 14 |

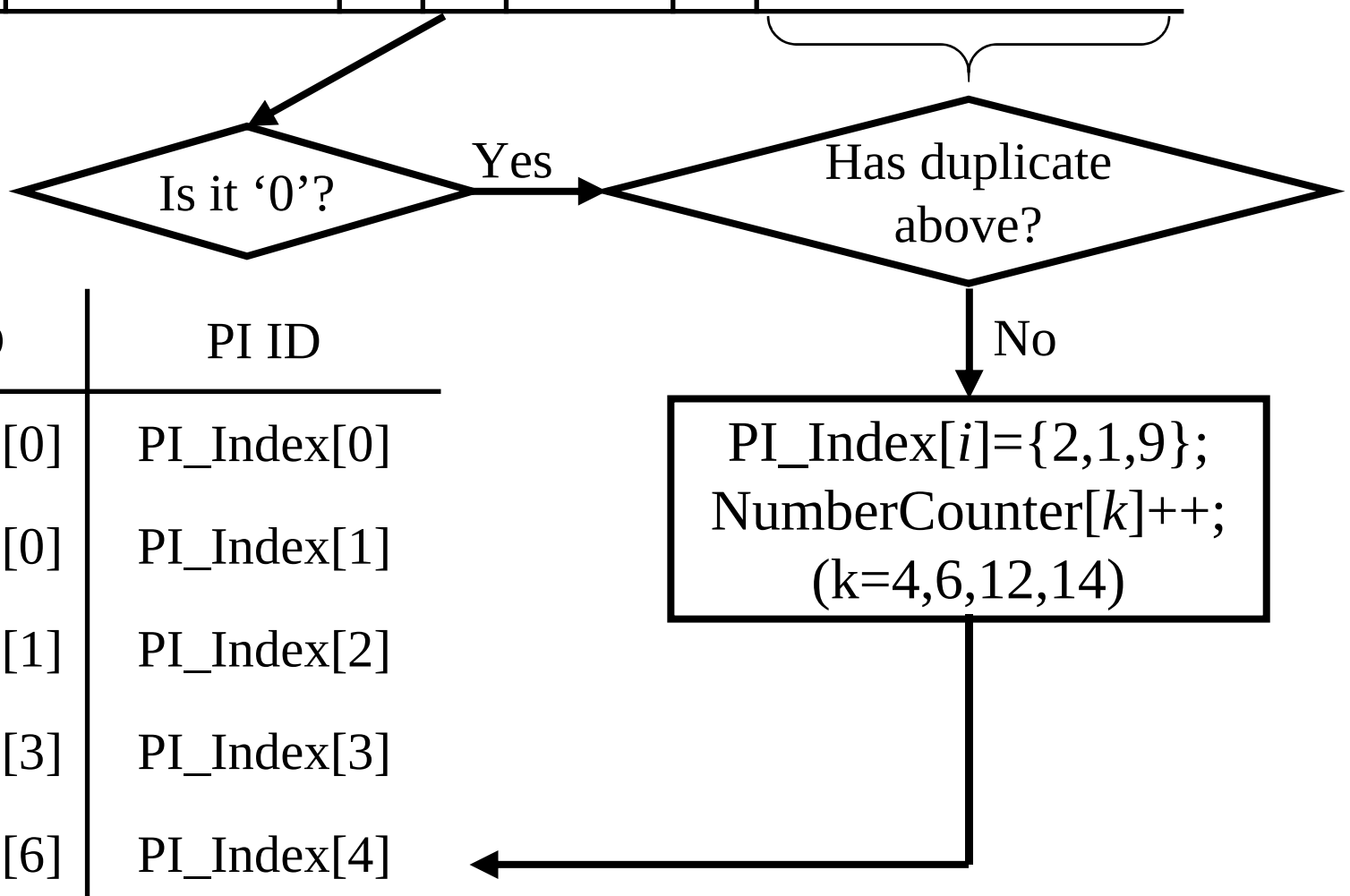

| PI | Column ID | PI ID |
|---|---|---|
| 0,1,4,5 | Column[2][0][0] | PI_Index[0] |
| 1,3,5,7 | Column[2][1][0] | PI_Index[1] |
| 1,3,9,11 | Column[2][1][1] | PI_Index[2] |
| 1,5,9,13 | Column[2][1][3] | PI_Index[3] |
| 4,5,6,7 | Column[2][1][6] | PI_Index[4] |
| 4,5,12,13 | Column[2][1][7] | PI_Index[5] |
| 4,6,12,14 | Column[2][1][9] | PI_Index[6] |

Fig 8. Addresses of unpaired PIs are recorded in the array PI_Index[][] (duplicates are omitted).

# Procedures of QM method (4)

PI Chart

| | | 0 | 1 | 3 | 4 | 5 | 6 | 7 | 9 | 11 | 12 | 13 | 14 |
|---|---|---|---|---|---|---|---|---|---|---|---|---|---|
| 0,1,4,5 | 0X0X | ● | ● | | ● | ● | | | | | | | |
| 1,3,5,7 | 0XX1 | | ● | ● | | ● | | ● | | | | | |
| 1,3,9,11 | X0X1 | | ● | ● | | | | | ● | ● | | | |
| 1,5,9,13 | XX01 | | ● | | | ● | | | ● | | | ● | |
| 4,5,6,7 | 01XX | | | | ● | ● | ● | ● | | | | | |
| 4,5,12,13 | X10X | | | | ● | ● | | | | | ● | ● | |
| 4,6,12,14 | X1X0 | | | | ● | | ● | | | | ● | | ● |

Fig 9. Produce the PI Chart. Columns of Don’t-Care minterms are removed

---

# Data structure and algorithm for implementation (4)

| *i* | 0 | 1 | 2 | 3 | 4 | 5 | 6 | 7 | 8 | 9 | 10 | 11 | 12 | 13 | 14 | 15 |
|---|---|---|---|---|---|---|---|---|---|---|---|---|---|---|---|---|
| NumberCounter[*i*] | 1 | 4 | 0 | 2 | 4 | 5 | 2 | 2 | 0 | 2 | 0 | 1 | 2 | 2 | 1 | 0 |

↓ Omit Don’t-Care minterms

| *i* | 0 | 1 | 2 | 3 | 4 | 5 | 6 | 7 | 8 | 9 | 10 | 11 | 12 | 13 | 14 | 15 |
|---|---|---|---|---|---|---|---|---|---|---|---|---|---|---|---|---|
| NumberCounter[*i*] | 0 | 0 | 0 | 0 | 4 | 5 | 2 | 0 | 0 | 2 | 0 | 1 | 2 | 2 | 1 | 0 |

Fig 10. Set the value of NumberCounter[*i*] to ‘0’ ( *i* is the Don’t-Care minterms).

## Procedures of QM method (5)

PI Chart

| | | | 4 | 5 | 6 | 9 | 11 | 12 | 13 | 14 |
|---|---|---|---|---|---|---|---|---|---|---|
| | 0,1,4,5 | 0X0X | ● | ● | | | | | | |
| | 1,3,5,7 | 0XX1 | | ● | | | | | | |
| ✓ | 1,3,9,11 | X0X1 | | | | ● | ● | | | |
| | 1,5,9,13 | XX01 | | ● | | ● | | | ● | |
| | 4,5,6,7 | 01XX | ● | ● | ● | | | | | |
| | 4,5,12,13 | X10X | ● | ● | | | | ● | ● | |
| ✓ | 4,6,12,14 | X1X0 | ● | | ● | | | ● | | ● |

Fig 11. Find the columns that contains only one mark. Pick the rows that cover this mark. The corresponding PIs will be EPIs.

---

## Data structure and algorithm for implementation (5)

| *i* | 0 | 1 | 2 | 3 | 4 | 5 | 6 | 7 | 8 | 9 | 10 | 11 | 12 | 13 | 14 | 15 |
|---|---|---|---|---|---|---|---|---|---|---|---|---|---|---|---|---|
| NumberCounter[*i*] | 0 | 0 | 0 | 0 | 4 | 5 | 2 | 0 | 0 | 2 | 0 | 1 | 2 | 2 | 1 | 0 |

1,3,9,11

4,6,12,14

| PI | Column ID | PI ID | EPI ID |
|---|---|---|---|
| 0,1,4,5 | Column[2][0][0] | PI_Index[0] | |
| 1,3,5,7 | Column[2][1][0] | PI_Index[1] | |
| 1,3,9,11 | Column[2][1][1] | PI_Index[2] | EPI_Index[0] |
| 1,5,9,13 | Column[2][1][3] | PI_Index[3] | |
| 4,5,6,7 | Column[2][1][6] | PI_Index[4] | |
| 4,5,12,13 | Column[2][1][7] | PI_Index[5] | |
| 4,6,12,14 | Column[2][1][9] | PI_Index[6] | EPI_Index[1] |

| *i* | 0 | 1 | 2 | 3 | 4 | 5 | 6 | 7 | 8 | 9 | 10 | 11 | 12 | 13 | 14 | 15 |
|---|---|---|---|---|---|---|---|---|---|---|---|---|---|---|---|---|
| NumberCounter[*i*] | 0 | 0 | 0 | 0 | 0 | 5 | 0 | 0 | 0 | 0 | 0 | 0 | 0 | 2 | 0 | 0 |

Fig 12. Find the minterms whose NumberCounter value is ‘1’. The addresses of the corresponding PIs that contain these minterms are recorded in the array EPI_Index[][].

## Procedures of QM method (6)

Reduced PI Chart

| | | 5 | 13 |
|---|---|---|---|
| 0,1,4,5 | 0X0X | ● | |
| 1,3,5,7 | 0XX1 | ● | |
| ✓ 1,5,9,13 | XX01 | ● | ● |
| 4,5,6,7 | 01XX | ● | |
| 4,5,12,13 | X10X | ● | ● |

**F(A,B,C,D) = B’D+BD’+C’D**

Fig 13. Produce the Reduced PI Chart. Select the least possible number of PIs that cover all the minterms

---

## Data structure and algorithm for implementation (6)

NumberOfRemainingMT=2;

ReducedPIChart_X[]={5,13};

NumberOfRemainingPI=5;

ReducedPIChart_Y[][]={PI_Index[0], PI_Index[1], PI_Index[3], PI_Index[4], PI_Index[5]};

The Reduced PI Chart will be represented by the two-dementional array ReducedPIChart[][].

Both the two PIs ‘1,5,9,13’ and ‘4,5,12,13’ can be the third EPI.
EPI_Index[2] will be ‘1,5,9,13’, which is found earlier in the recursive for loops.

| | Column ID | EPI ID |
|---|---|---|
| 1,3,9,11 | Column[2][1][1] | EPI_Index[0] |
| 1,5,9,13 | Column[2][1][3] | EPI_Index[2] |
| 4,6,12,14 | Column[2][1][9] | EPI_Index[1] |

**F(A,B,C,D) = B’D+BD’+C’D**

Fig 14. Produce the Reduced PI Chart and find the rest EPIs

## 3. C Codes for implementation of QM method

```
#include <stdio.h>
#include <stdlib.h>
#include <math.h>

int i,j,temp,NumberOfVariable,NumberOfAllMinterm,NumberOfDontCare,NumberOfEPI=0,
NumberOfRemainingMT,NumberOfRemainingPI,NumberOfPI=0,PotEPINo=0,NumberOfPossibleEPI=
1,MinimumNo=0,Groupable=1;
int *MintermIndicesDecimal, *MintermIndicesDecimal_DontCare, **Minterm_Binary,
****Column, **PI_Index,
**EPI_Index,*NumberCounter,*ReducedPIChart_X,**ReducedPIChart_Y,**ReducedPIChart,
*For, **Potential_EPI,*NoOfPIForEPI;

void DecimalToBinary();
int OneCounter(int *binary, int NumberOfDigit);
int Combination(int n, int ColumnNo, int k);
int IsPowerOfTwo(int n);
int IsDontCare(int MT);
void ShowResult();
void Recursion_For_Loop(int m);

int main()
{
     int k,l,m,n,x,y,LogicProbe;

/**********Preparation. Collect the information for the boolean
expression**********/
     printf("Please provide the information for the sum of minterms.\n\nHow many
variables does it contain?\n");
     scanf("%d",&NumberOfVariable);
     while(NumberOfVariable<=0)
     {
             printf("The number of variables should be greater than 0, please enter
again:\n\n");
             printf("Please provide the information for the sum of minterms.\n\nHow
many variables does it contain?\n");
             scanf("%d",&NumberOfVariable);
     }

     printf("How many minterms (including Don't-Care minterms) does it contain?\n");
     scanf("%d",&NumberOfAllMinterm);
     while(NumberOfAllMinterm>pow(2,NumberOfVariable) || NumberOfAllMinterm<=0)
     {
             printf("The number of minterms cannot be greater than 2^%d nor smaller
than 1, please enter again:\n",NumberOfVariable);
             printf("How many minterms (including Don't-Care minterms) does it
contain?\n");
             scanf("%d",&NumberOfAllMinterm);
     }

     printf("How many Don't-Care minterms does it contain?\n");
     scanf("%d",&NumberOfDontCare);
     while(NumberOfDontCare>=NumberOfAllMinterm || NumberOfDontCare<0)
     {
             printf("The number of Don't-Care minterms cannot be greater than the
```

```
number of all minterms nor smaller than 0, please enter again:\n");
            printf("How many Don't-Care minterms does it contain?\n");
            scanf("%d",&NumberOfDontCare);
    }

    MintermIndicesDecimal=(int *)malloc(NumberOfAllMinterm*sizeof(int));
    /* Record the decimal indices representing each minterm */
    MintermIndicesDecimal_DontCare=(int *)malloc(NumberOfDontCare*sizeof(int));
    /* Record the decimal indices representing Don't-Care minterms */

    for(i=0;i<NumberOfAllMinterm;i++)
    {
            if(i==0)
                    printf("Please enter the decimal index of the 1st minterm(in
ascending order):");
            else if(i==1)
                    printf("Please enter the decimal index of the 2nd minterm(in
ascending order):");
            else if(i==2)
                    printf("Please enter the decimal index of the 3rd minterm(in
ascending order):");
            else
                    printf("Please enter the decimal index of the %dth minterm(in
ascending order):",i+1);

            scanf("%d",&MintermIndicesDecimal[i]);
            if(i!=0 && MintermIndicesDecimal[i]<=MintermIndicesDecimal[i-1])
            {
                    printf("The numbers are not in ascending order, please re-enter
all the indices again.\n\n");
                    i=-1;
            }
            else if(MintermIndicesDecimal[i]>=pow(2,NumberOfVariable))
            {
                    printf("\nThe number should be smaller than %d, please re-enter
all the indices again.\n\n",pow(2,NumberOfVariable));
                    i=-1;
            }
    }


    if(NumberOfDontCare!=0)
    {
            printf("\n\nWhich of them are Don't-Care terms?\n\n");
            for(i=0;i<NumberOfDontCare;i++)
            {
                    if(i==0)
                            printf("Please enter the decimal index of the 1st Don't-
Care minterm (in ascending order):");
                    else if(i==1)
                            printf("Please enter the decimal index of the 2nd Don't-
Care minterm (in ascending order):");
                    else if(i==2)
                            printf("Please enter the decimal index of the 3rd Don't-
Care minterm (in ascending order):");
                    else
```

```
			printf("Please enter the decimal index of the %dth Don't-Care minterm (in ascending order):",i+1);
		scanf("%d",&MintermIndicesDecimal_DontCare[i]);
		if(i!=0 && MintermIndicesDecimal_DontCare[i]<=MintermIndicesDecimal_DontCare[i-1])
		{
			printf("The numbers are not in ascending order, please re-enter all the indices again.\n\n");
			i=-1;
		}
		else if(MintermIndicesDecimal[i]>=pow(2,NumberOfVariable))
		{
			printf("\nThe number should be smaller than %d, please re-enter all the indices again.\n\n",pow(2,NumberOfVariable));
			i=-1;
		}
	}
}

/***********Transform the decimal indices into Binary format and obtain relative information.***********/
Minterm_Binary=(int **)malloc(NumberOfAllMinterm*sizeof(int*));
for(i=0;i<=NumberOfAllMinterm;i++)
{
	Minterm_Binary[i]=(int *)malloc((NumberOfVariable+4)*sizeof(int));
}
DecimalToBinary();
for(i=0;i<NumberOfAllMinterm;i++)
{

Minterm_Binary[i][NumberOfVariable]=OneCounter(Minterm_Binary[i],NumberOfVariable);
	Minterm_Binary[i][NumberOfVariable+1]=0;
		/* '0' means it hasn't been grouped, '1' means it has been grouped with other terms */
	Minterm_Binary[i][NumberOfVariable+2]=MintermIndicesDecimal[i];
/* this is its original minterm */
	Minterm_Binary[i][NumberOfVariable+3]=MintermIndicesDecimal[i];
/* this is all the minterms it consists of */
}

/***********Prepare the first column for grouping***********/
Column=(int ****)malloc((NumberOfVariable+1)*sizeof(int***));
for(i=0;i<NumberOfVariable+1;i++)
{
	Column[i]=(int ***)malloc((NumberOfVariable+1-i)*sizeof(int**));
			/* Column[i] contains all the terms in the (i+1)th column */
}
for(i=0;i<NumberOfVariable+1;i++)
{
	for(j=0;j<NumberOfVariable+1-i;j++)
	{
		Column[i][j]=(int **)malloc(Combination(NumberOfVariable,i,j)*sizeof(int*));		/* Column[i][j] contains all the terms with j '1's in their binary indices in the (i+1)th column */
```

```
                    for(k=0;k<Combination(NumberOfVariable,i,j);k++)
                    {
                            Column[i][j][k]=NULL;
                                                        /* Column[i][j][k]
represents a term with in the j'1's in their binary indices in the (i+1)th column
*/
                    }
            }
    }
    for(i=0;i<NumberOfVariable+1;i++)
    {
            for(j=0,k=0;j<NumberOfAllMinterm;j++)
            {
                    if(Minterm_Binary[j][NumberOfVariable]==i)
                    {
                            Column[0][i][k++]=Minterm_Binary[j];
                    /* Prepare the first grouping column */
                    }
            }
    }

/***********Perform the grouping***********/
    for(i=0;i<NumberOfVariable+1;i++)
    {
            if(Groupable)
            {
                    Groupable=0;
                            for(j=0;j<NumberOfVariable-i;j++)
                            {
                                    int p,position;
                                    m=0;
                                    for(k=0;k<Combination(NumberOfVariable,i,j);k++)
                                            if(Column[i][j][k]!=NULL)
                                            {

    for(l=0;l<Combination(NumberOfVariable,i,j+1);l++)
                                                    {
                                                            if(Column[i][j+1][l]!=NULL
&& Column[i][j+1][l][NumberOfVariable+2+i]>Column[i][j][k][NumberOfVariable+2+i] &&
IsPowerOfTwo(Column[i][j+1][l][NumberOfVariable+2+i]-
Column[i][j][k][NumberOfVariable+2+i]))
                                                            {
                                                                    LogicProbe=0-i;
/*This LogicProbe is used to check whether this two terms has the same positions of
'-'(which is represented by '2')*/
                                                                    for(n=1;n<=i;n++)

    for(p=1;p<=i;p++)

    if(Column[i][j+1][l][NumberOfVariable+1+n]==Column[i][j][k][NumberOfVariable+1+
p])
                                                                                    {

    LogicProbe++;
                                                                                    }
                                                                    if(LogicProbe==0)
```

```
                                                                    {
    Groupable=1;
    Column[i][j][k][NumberOfVariable+1]=1;
    Column[i][j+1][l][NumberOfVariable+1]=1;
    Column[i+1][j][m]=(int *)malloc((NumberOfVariable+4+i+pow(2,i+1))*sizeof(int));
    for(n=0;n<=NumberOfVariable+1+i;n++)
                                                                          {
    Column[i+1][j][m][n]=Column[i][j][k][n];
                                                                          }

    Column[i+1][j][m][NumberOfVariable+3+i]=Column[i][j][k][NumberOfVariable+2+i];

    for(n=NumberOfVariable+4+i;n<NumberOfVariable+4+i+pow(2,i+1);n++)
    Column[i+1][j][m][n]=0;

    position=log((Column[i][j+1][l][NumberOfVariable+2+i]-
Column[i][j][k][NumberOfVariable+2+i]))/log(2);

    Column[i+1][j][m][NumberOfVariable-1-position]=2;
    Column[i+1][j][m][NumberOfVariable+1]=0;
    Column[i+1][j][m][NumberOfVariable+2+i]=position;

    for(p=0;p<pow(2,i);p++)
                                                                          {
    Column[i+1][j][m][NumberOfVariable+4+i+p]=Column[i][j][k][NumberOfVariable+3+i+
p];
                                                                          }
    for(p=pow(2,i);p<pow(2,i+1);p++)
                                                                          {
    Column[i+1][j][m][NumberOfVariable+4+i+p]=Column[i][j+1][l][NumberOfVariable+3+
i+p-(int)pow(2,i)];
                                                                          }
                                                                          m++;
                                                                          }
                                                              }
                                                        }
                                            }
                                      }
                          }
```

```
    }

/***********NumberCounter count how many times each decimal index occurs***********/
    NumberCounter=(int *)malloc(pow(2,NumberOfVariable)*sizeof(int));
    for(i=0;i<pow(2,NumberOfVariable);i++)
            NumberCounter[i]=0;

/***********Record the Prime Implicants(duplicates will be removed)***********/
    PI_Index=(int **)malloc(NumberOfAllMinterm*sizeof(int*));
    for(i=0;i<NumberOfAllMinterm;i++)
    {
            PI_Index[i]=(int *)malloc(3*sizeof(int));
    }

    for(i=0;i<NumberOfVariable+1;i++)
            for(j=0;j<NumberOfVariable+1-i;j++)
                    for(k=0;k<Combination(NumberOfVariable,i,j);k++)
                    {
                            if(Column[i][j][k]!=NULL &&
Column[i][j][k][NumberOfVariable+1]==0)
                            {
                                    LogicProbe=0-pow(2,i); /*LogicProbe is used to
check whether this PI is a duplicate*/
                                    for(l=k-1;l>=0;l--)
                                    if(LogicProbe!=0)
                                    {
                                            LogicProbe=0-pow(2,i);
                                            for(m=0;m<pow(2,i);m++)
                                                    for(n=0;n<pow(2,i);n++)

    if(Column[i][j][l][NumberOfVariable+3+i+m]==Column[i][j][k][NumberOfVariable+3+
i+n])
                                                    {
                                                            LogicProbe++;
                                                    }
                                    }
                                    if(LogicProbe!=0)
                                    {
                                            PI_Index[NumberOfPI][0]=i;
                                            PI_Index[NumberOfPI][1]=j;
                                            PI_Index[NumberOfPI][2]=k;
                                            NumberOfPI++;
                                            for(l=0;l<pow(2,i);l++)
                                            {

    NumberCounter[Column[i][j][k][NumberOfVariable+3+i+l]]++;
                                            }
                                    }
                            }
                    }
/***********Remove the DontCare minterms***********/
    for(i=0;i<NumberOfDontCare;i++)
            NumberCounter[MintermIndicesDecimal_DontCare[i]]=0;

    EPI_Index=(int **)malloc(NumberOfAllMinterm*sizeof(int*));
/***********In the PI Chart, find the minterms which only occurs once, and select
```

```
the PIs which contain these minterms as EPIs and record them. Then set NumberCounter
of this minterms to 0***********/
    for(i=0;i<pow(2,NumberOfVariable);i++)
        if(NumberCounter[i]==1)
            for(j=0;j<NumberOfPI;j++)
                for(k=0;k<pow(2,PI_Index[j][0]);k++)
                {

    if(Column[PI_Index[j][0]][PI_Index[j][1]][PI_Index[j][2]][NumberOfVariable+3+PI
_Index[j][0]+k]==i)
                    {
                        EPI_Index[NumberOfEPI]=PI_Index[j];
                        for(l=0;l<pow(2,PI_Index[j][0]);l++)

    NumberCounter[Column[PI_Index[j][0]][PI_Index[j][1]][PI_Index[j][2]][NumberOfVa
riable+3+PI_Index[j][0]+l]]=0;
                        NumberOfEPI++;
                        k=pow(2,PI_Index[j][0]);
                    }
                }

/***********Make the Reduced PI Chart***********/
    NumberOfRemainingMT=0;
    for(i=0;i<pow(2,NumberOfVariable);i++)
        if(NumberCounter[i]!=0)
            NumberOfRemainingMT++;

    ReducedPIChart_X=(int *)malloc(NumberOfRemainingMT*sizeof(int));
    for(i=0;i<NumberOfRemainingMT;i++)
        ReducedPIChart_X[i]=-1;

    ReducedPIChart_Y=(int **)malloc(NumberOfPI*sizeof(int*));
    for(i=0;i<NumberOfPI;i++)
        ReducedPIChart_Y[i]=NULL;

    ReducedPIChart=(int **)malloc(NumberOfRemainingMT*sizeof(int*));

/***********This is the First Row, consist of the remaining minterms decimal
indices***********/
    for(i=0,j=0;j<pow(2,NumberOfVariable);j++)
        if(NumberCounter[j]!=0)
        {
            ReducedPIChart_X[i]=j;
            i++;
        }

/***********This is the First Column, consist of the remaining PIs***********/
    NumberOfRemainingPI=0;
    for(i=0;i<NumberOfPI;i++)
        for(j=0;j<pow(2,PI_Index[i][0]);j++)
        {

    if(NumberCounter[Column[PI_Index[i][0]][PI_Index[i][1]][PI_Index[i][2]][NumberO
fVariable+3+PI_Index[i][0]+j]]!=0)
            {
                j=pow(2,PI_Index[i][0]);
```

```c
                    ReducedPIChart_Y[NumberOfRemainingPI]=PI_Index[i];
                    NumberOfRemainingPI++;
                }
            }

/***********ReducedPIChart[i][j] represent the information of Reduced PI Chart('1'
means picked, '0' means unpicked)***********/
    if(NumberOfRemainingPI!=0)
    {
            for(i=0;i<NumberOfRemainingMT;i++)
            ReducedPIChart[i]=(int *)malloc(NumberOfRemainingPI*sizeof(int));

            for(i=0;i<NumberOfRemainingMT;i++)
                for(j=0;j<NumberOfRemainingPI;j++)
                    ReducedPIChart[i][j]=0;

            for(i=0;i<NumberOfRemainingPI;i++)
                for(j=0;j<pow(2,ReducedPIChart_Y[i][0]);j++)
                    for(k=0;k<NumberOfRemainingMT;k++)

    if(Column[ReducedPIChart_Y[i][0]][ReducedPIChart_Y[i][1]][ReducedPIChart_Y[i][2
]][NumberOfVariable+3+ReducedPIChart_Y[i][0]+j]==ReducedPIChart_X[k])
                                    {
                                            ReducedPIChart[k][i]=1;
                                    }

/***********Select the EPIs from the Reduced PI Chart***********/
            For=(int *)malloc(NumberOfRemainingMT*sizeof(int));  /* For[i] will be
used in the function 'Recursion_For_Loop(int m)' */
            for(i=0;i<NumberOfRemainingMT;i++)
            {
                For[i]=-1;
            }

            for(i=0;i<NumberOfRemainingMT;i++)

    NumberOfPossibleEPI=NumberOfPossibleEPI*NumberCounter[ReducedPIChart_X[i]];

            Potential_EPI=(int **)malloc(NumberOfPossibleEPI*sizeof(int*));

            for(i=0;i<NumberOfPossibleEPI;i++)
            {
                Potential_EPI[i]=(int *)malloc(NumberOfRemainingMT*sizeof(int));
            }
            Recursion_For_Loop(NumberOfRemainingMT-1);

            NoOfPIForEPI=(int *)malloc(NumberOfPossibleEPI*sizeof(int)); /*
NoOfPIForEPI[i] will count how many PIs are in each combination that covers all
minterms */
            for(i=0;i<NumberOfPossibleEPI;i++)
                NoOfPIForEPI[i]=0;

            for(i=0;i<NumberOfPossibleEPI;i++)
                for(j=0;j<NumberOfRemainingMT;j++)
                    if(Potential_EPI[i][j]!=-1)
                    {
```

```c
                                NoOfPIForEPI[i]++;
                                for(k=j+1;k<NumberOfRemainingMT;k++)

    if(Potential_EPI[i][k]==Potential_EPI[i][j])
                                            Potential_EPI[i][k]=-1;
                        }

/**********Find the combination which require the least number of PIs to cover all minterms**********/
        for(i=1;i<NumberOfPossibleEPI;i++)
                if(NoOfPIForEPI[i]<NoOfPIForEPI[MinimumNo])
                        MinimumNo=i;

        for(i=0;i<NumberOfRemainingMT;i++)
                if(Potential_EPI[MinimumNo][i]!=-1)

    EPI_Index[NumberOfEPI++]=ReducedPIChart_Y[Potential_EPI[MinimumNo][i]];

/**********Print the final result of minimal SOP expression**********/
        printf("\nThe simplified SOP expression is:\n\n");
        printf("\n      ");
        for(x=0;x<NumberOfEPI;x++)
        {
                for(y=0;y<NumberOfVariable;y++)
                {

    if(Column[EPI_Index[x][0]][EPI_Index[x][1]][EPI_Index[x][2]][y]==1)
                                printf("%c",65+y);
                        else
if(Column[EPI_Index[x][0]][EPI_Index[x][1]][EPI_Index[x][2]][y]==0)
                                printf("%c'",65+y);
                }
                if(x<NumberOfEPI-1)
                        printf("+");
        }
        printf("\n\nPress any key to exit...");
        scanf("%d",&i);
        return 0;
    }
    else
    {
        printf("\n\nThe simplified SOP expression is:\n\n");
        printf("\n      ");
        for(x=0;x<NumberOfEPI;x++)
        {
                for(y=0;y<NumberOfVariable;y++)
                {

    if(Column[EPI_Index[x][0]][EPI_Index[x][1]][EPI_Index[x][2]][y]==1)
                                printf("%c",65+y);
                        else
if(Column[EPI_Index[x][0]][EPI_Index[x][1]][EPI_Index[x][2]][y]==0)
                                printf("%c'",65+y);
                }
                if(x<NumberOfEPI-1)
                        printf("+");
```

```c
		}
		printf("\n\nPress any key to exit...");
		scanf("%d",&i);
		return 0;
	}
}

int IsDontCare(int MT)
{
	int i;
		for(i=0;i<NumberOfDontCare;i++)
			if(MT==MintermIndicesDecimal_DontCare[i])
				return 1;
		return 0;
}

void DecimalToBinary()
{
	int i,j,dividend;
	for(i=0;i<NumberOfAllMinterm;i++)
	{
		dividend=MintermIndicesDecimal[i];
		for(j=NumberOfVariable-1;j>=0;j--)
		{
			Minterm_Binary[i][j]=dividend%2;
			dividend=dividend/2;
		}
	}
}

int OneCounter(int *binary, int NumberOfDigit)
{
	int i,count=0;
	for(i=0;i<=NumberOfDigit-1;i++)
	{
		if(binary[i]==1)
			count++;
	}
	return count;
}

int Combination(int n, int ColumnNo, int k)
{
	int Comb,i,NtoK=1,Kto1=1;
	for(i=n;i>=n-k+1-ColumnNo;i--)
	{
		NtoK=i*NtoK;
	}
	for(i=k;i>=1;i--)
	{
		Kto1=i*Kto1;
	}
	Comb=NtoK/Kto1;
	return Comb;
}
```

```
int IsPowerOfTwo(int n)
{
     return(floor(log(n)/log(2))==(log(n)/log(2)));
}

void Recursion_For_Loop(int m)
{
     int n=m;
     for(For[n]=0;For[n]<NumberOfRemainingPI;For[n]++)
     {
                    if(ReducedPIChart[NumberOfRemainingMT-1-n][For[n]])
                    {
                         if(n>0)
                         {
                              m=n;
                              m--;
                              Recursion_For_Loop(m);
                         }
                         else if(n==0)
                         {
                              for(i=0;i<NumberOfRemainingMT;i++)
                              {

     Potential_EPI[PotEPINo][i]=For[NumberOfRemainingMT-1-i];
                              }
                              PotEPINo++;
                         }
                    }
     }
}
```

## 4. Conclusion

The aim of this article is to introduce a QM simulation code based on C program that is able to deal with any number of variables and takes the Don't-Care condition into consideration, which is not easy to obtain online. Each step of QM method is illustrated in parallel with the corresponding data structure and algorithm used in the program to make it easier to understand. This program is also considered as a good example for practicing program implementation of methods. Any suggestion to improve this program will be appreciated.